\documentclass[a4paper,10pt,oneside,onecolumn,preprint]{elsarticle}
\usepackage{amsmath}
\usepackage{amssymb}
\usepackage{latexsym}
\usepackage{rotating}
\usepackage{cite}
\usepackage{pifont}
\usepackage{multirow}
\usepackage[T1]{fontenc}
\usepackage{float}
\usepackage{caption}
\numberwithin{equation}{section}
\renewcommand{\figurename}{Fig.}
\begin{document}
\newcommand{\promille}{\%\symbol{'30} }
\newcommand{\ep}{\varepsilon}
\newcommand{\al}{\alpha_{ik}}
\newcommand{\be}{\beta_{ik}}

\title{The critical energy of the Ising double chain - the explanation for melting}
\author{Klaus Mika\fnref{}}
\ead{klaus.mika.juelich@gmail.com}
\address{Institut f\"ur Festk\"orperforschung, Forschungszentrum J\"ulich, 52425 J\"ulich, Germany}

\begin{abstract}
The Ising double chain has been overlooked for more than 80 years, since
one was mostly interested in phase transitions. We consider two types of degeneracy, spin occupation degeneracy and "bond" occupation degeneracy. We
show that bond occupation on "open nets" is of particular importance for the
understanding of melting. The number of open nets on a double chain is
determined analytically. If only one bond site of a net is occupied, this already
produces a definite number of bonds on sites, bonds that do not belong to the net. All
nets on a given chain of length $n$ then yield a total number $i^{(n)}_{\rm tot}$ 
 of  bonds,  allowing to calculate a characteristic energy $u^{(n)}_{ch}$, an  approximation of the critical energy  $u_c$  of the infinite double chain. We finally present an explanation for the critical energy $u^{\rm KG}_c$   of the kagomé lattice in terms of the critical energies  $u^{\rm TR}_c$ and $u^{\rm HC}_c$  of the triangular and honeycomb lattices.
 \end{abstract}


\maketitle
\section{Introduction}

J. W. Gibbs made a fundamental statement at the beginning of his work on 
heterogeneous equilibria around 1875, which he however never applied 
in his later work \citep{gib}:
 {\sl The comprehension of the laws which govern any material 
system is greatly facilitated by considering the entropy and energy of the system 
in the various states of which it is capable.} It is possible that the reason for this 
statement were the Carnot cycles, which have always been treated in 
variables leading to four hyperbolas instead of a rectangle. Also, at that time, 
microcanonical description was not yet known. 

\begin{figure}[h]
\centering\includegraphics[width=120mm]{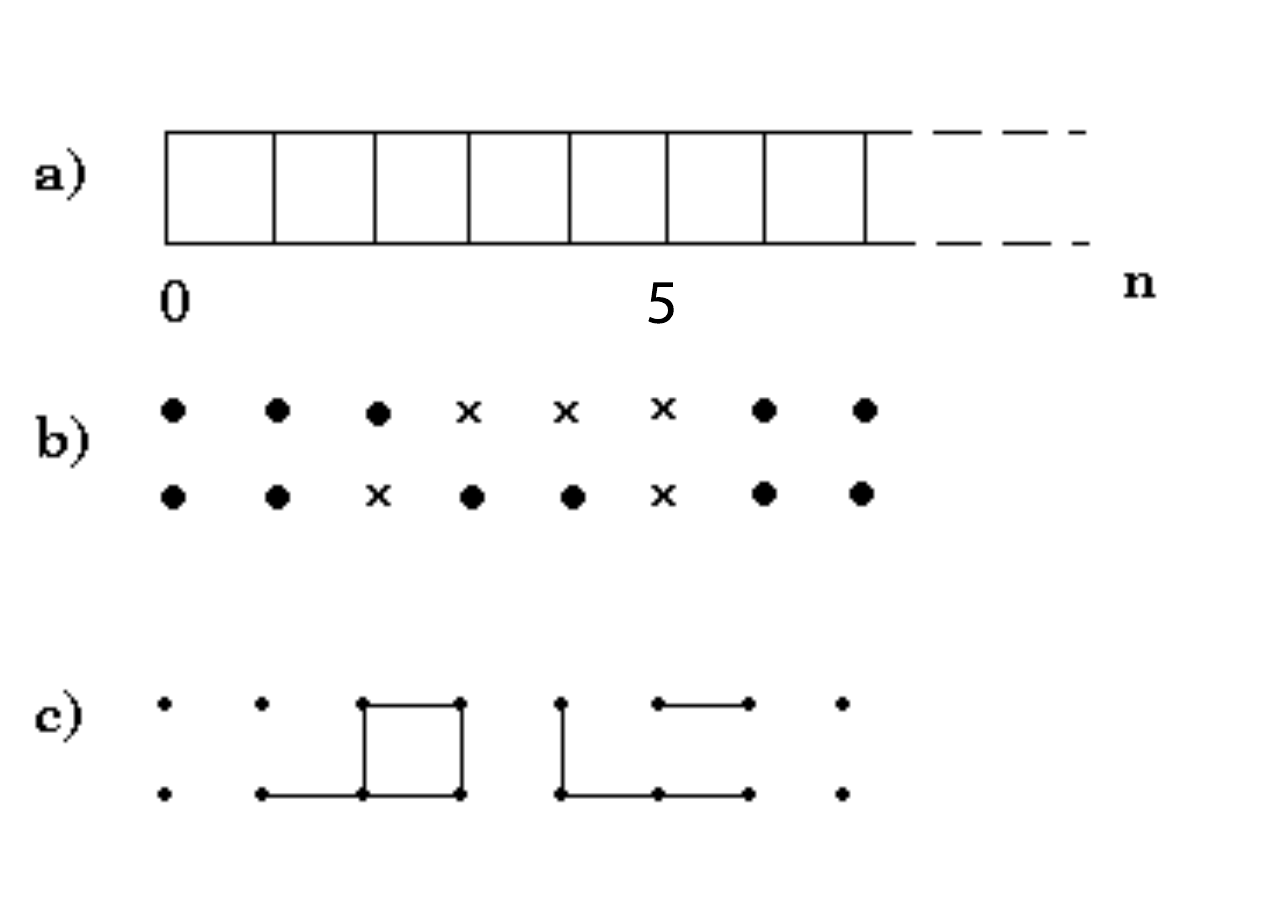} 
\caption{\label{mika-fig1} 
a)  Ising double chain of length $n=7$ with $N=16$ spins ($i.e.$ lattice points) and  $L=22$ bond sites ($i.e.$ cell sides). 
 b) Example of a possible arrangement of spins on the  double chain (dots and crosses correspond to spin values of +1 and -1 respectiveley). 
 c) Bonds on the double chain shown in Fig. 1b.}
\end{figure}

In the present work, we will consider the Ising double chain. A double chain of length $n$ has the 
form shown in \figurename~\ref{mika-fig1}a. Each lattice point is characterized (\figurename~\ref{mika-fig1}b) by spin values +1 and -1.

The two-dimensional Ising model  was solved by L. Onsager \citep{ons} and B. Kaufman \citep{kau}. They used a canonical description: From the partition function, they obtained the entropy $S$ as a function of temperature $T$ and the critical temperature $T_c$. The special case of the linear chain, solved by E. Ising \citep{isi}, did not lead to a critical temperature. The 
double chain has rarely been considered, perhaps because it has no critical temperature and therefore no phase transition, and as such is believed uninteresting. We will show that it actually possesses some interesting properties.

\begin{figure}[h]
\centering\includegraphics[width=120mm]{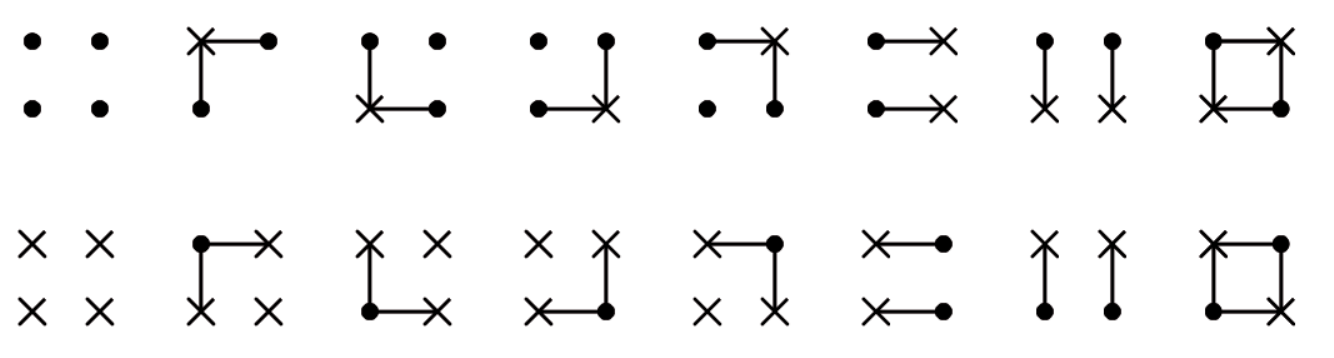} 
\caption{\label{mika-fig2} The 16 and 8 possible spin and bond arrangements respectively on a double chain of length $n=1$.}
\end{figure}

To describe the Ising double chain, we introduce the concept of "bonds", that 
connect neighbouring lattice points with {\em opposite} spins (\figurename~\ref{mika-fig1}c). For $n = 1$, we 
have already 16 different spin arrangements, but 8 different bond arrangements (\figurename~\ref{mika-fig2}). 
The double chain does not allow the free occupation of its bond sites with bonds. This has to be taken into account when placing bonds on the chain; the bond arrangement has to obey the {\em constraint} that each cell must have an even number of bonds.

We will not use a canonical description, $i.e.$ we ignore the temperature, nor do we follow a complete microcanonical approach; instead, we limit ourselves to two special types of 
microcanonical description: spin occupation degeneracy, and in particular  $bond$ 
$occupation$ degeneracy. While spin occupation always leads to long-range order (mean-field theory), we shall see that bond occupation always produces disorder. By applying bond occupation to the double chain, we will obtain the entropy $S$ as a function of the energy $u$ and will gain access to the critical energy $u_c$.  At  $u = u_c$,   the bond entropy $S$ is 0 and grows to $\ln2$  as  $u \to 0$   \citep{mik}. 

We will evaluate a special type of bond arrangement in order to obtain {\em directly} the characteristic energy  $u^{(n)}_{ch}$  of a double chain of finite length $n$. With increasing $n$,  $u^{(n)}_{ch}$ converges rapidly to the critical energy $u_c$ of the infinite chain.

\section{Two different types of occupation}
We use the following normalization to denote the internal energy  $u$  of the Ising systems:  $u=1$ for the ferromagnetic ground state and  $u = 0$ for the state of total disorder.
Therefore the entropy $S(u)$ per particle satisfies $S(1) = 0$ and has its
maximum value $S(0) = \ln 2$ at $u=0$. Each of the $N$ spins can take
the values +1 and -1.  We will further differentiate the expressions for the entropies by an upper index, $S^{(0)}$ and $S^{(1)}$, to designate
the specific form taken near $u =0$ and $u=1$ respectively. The internal energy $u$ is given by the number
$i$ of bonds of neighboring spins with opposite sign 
\begin{equation} 
u = 1 - 2i/L, \label{eqn_uofn}
\end{equation} 
where $L$ is the total number of bond sites. Hence $u = 0$ for $i = L/2$.

\subsection{Spin occupation}
\begin{figure}[h]
\centering\includegraphics[width=120mm]{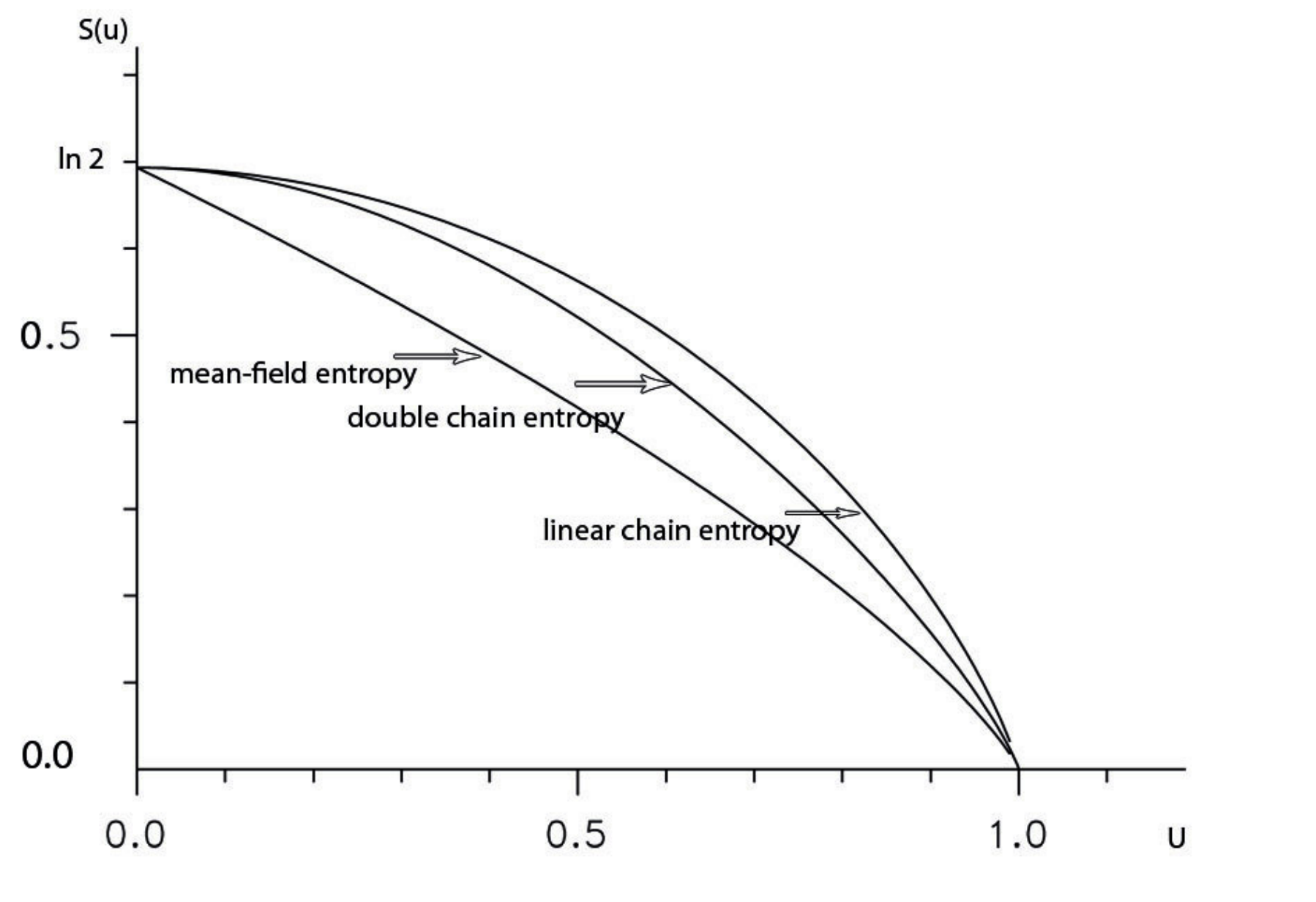} 
\caption{\label{mika-fig3} 
Lower curve: entropy $S_s(u)$ of the mean-field theory (Eq. \ref{eqn_Smf}). At $u=0$ we 
have a none-zero slope of $S_s(u)$ leading to a cut-off tangent. Upper curve: entropy of the linear chain (Eq. \ref{eqn_Slc}). 
Central curve: entropy of the double chain (Eq. \ref{eqn_Sdc}).}
\end{figure}

If $k$ spin sites are occupied by spins (-1), the degeneracy is $g(k) = \binom{N}{k}$,
and the spin entropy 
\begin{equation}
S_s(k) = \frac{1}{N} \ln g(k).
\end{equation}
 For large $N$, we have
\begin{equation}
S_s(k) = -\frac{k}{N} \ln \frac{k}{N} - \left( 1 - \frac{k}{N} \right) \ln \left( 1 - \frac{k}{N} \right).
\end{equation}
Let $p = k/N$ be the probability to find a spin  (-1). Then the probability
to find a bond site occupied is $w = 2 pq$, where $q = 1 - p$. The total 
number of bonds is $i = Lw$, and with $u = 1 - 2i/L = 1 - 4 pq$ we obtain
$p = (1 - \sqrt{u})/2$. The entropy $S_s(u)$  therefore is
\begin{equation}
S_s(u) = -\frac{1}{2} (1 - \sqrt{u}) \ln \frac{1}{2} (1 - \sqrt{u})
- \frac{1}{2} (1 + \sqrt{u}) \ln \frac{1}{2} (1 + \sqrt{u}),  \label{eqn_Smf}
\end{equation}
{\sl i.e.} the mean-field entropy, valid for all Ising systems (\figurename~\ref{mika-fig3}).

Near $u = 0$ we have 
\begin{equation}
S_s^{(0)}(u) = \ln 2 - \frac{1}{2} u,
\end{equation} 
{\sl i.e.} a finite slope at $u = 0$,
the mean-field $T_c$, which is a cut-off tangent and has no physical meaning,
except for the infinite-dimensional Ising system, where $S_s(u)$ is the
exact entropy. At $u = 1 - \ep$ ($\ep \ll 1$), we have  
\begin{equation}
S_s^{(1)}(u) = - \frac{\ep}{4} \ln \frac{\ep}{4} +\frac{\ep}{4} +O(\ep^2 \ln \ep).
\end{equation}
\noindent This result holds for all Ising systems with two and higher dimensions.

\subsection{Bond occupation}
The linear chain allows 
the free occupation of bond sites with bonds, $i.e.$ there are no constraints \citep{isi}. Let $i = k$ bonds be occupied and $u = 1 -
2k/L$, then $S_b(u) = S(u)$ and
\begin{equation}
S(u) = -\frac{1}{2} (1 - u) \ln \frac{1}{2} (1 - u) -\frac{1}{2}(1 + u)
\ln \frac{1}{2} (1 + u).  \label{eqn_Slc}
\end{equation}

\noindent Near $u = 0$ we have 
\begin{equation}
S^{(0)}(u) = \ln 2 - \frac{1}{2} u^2 .
\end{equation}

As bond sites on the linear chain can be occupied freely, no additional energy can occur
and the entropy is the maximum possible entropy of all Ising
systems. The entropy of spin occupation is the minimum entropy, because it uses the maximum energy per occupation with one spin. (Here we consider only the ferromagnetic branch.) 

\section {The Ising double chain}

\subsection{Thermodynamic results}

The double chain of length $n$ has $N = 2n + 2$ spin sites and $L = 3n + 1$
bond sites (see \figurename~1a for $n=7$). The internal energy $u$ and the entropy $S_{\rm dc}(u)$ can, for large $n$,  be written in terms of 
the parameter $x$, where $x$ is related to the temperature $T$ by $x = \tanh{K} = \tanh{J/kT}$ \citep{mik}:

\begin{eqnarray}
\lambda(x) 
 & = & \frac{1}{2}\Big[1 + x^2 + \sqrt{(1 - x^2)^2 + 4x^4}\Big], \\
u(x) & = & x \left\{ 1+\frac{1}{3\lambda(x)}(1-x^2) \left[ 1+\frac{5x^2-1}{\sqrt{(1-x^2)^2+4x^4}} \right] \right\}, \\
S_{\rm dc}(x) & = & -\frac{3}{4} u(x) \ln \frac{1+x}{1-x} +\ln 2 -\frac{3}{4} \ln (1-x^2) +
\frac{1}{2} \ln \lambda(x), \label{eqn_Sdc}
\end{eqnarray}
The mean-field (Eq. \ref{eqn_Smf}), the linear chain (Eq. \ref{eqn_Slc}) and the double chain (Eq. \ref{eqn_Sdc}) entropies are displayed for comparison in \figurename~\ref{mika-fig3}.

\noindent Near $x = 0$ (the state of total disorder) we have $\lambda(x) = 1 + x^4,
u(x) = x$ and $S_{\rm dc}^{(0)}(x) = \ln 2 - 3/4 ~ x^2$, or 
\begin{equation}
S_{\rm dc}^{(0)}(u) = \ln 2 -\frac{1}{2} \frac{L}{N} u^2. \label{eqn_Su0}
\end{equation}
\noindent This result holds for all Ising systems \citep{mik}.

At $x = 1 - \ep$ (the ferromagnetic ground state) we have $\lambda(x) = 2 - 3\ep + 2\ep^2$, $u(x) = 1 - 1/3 ~ \ep^2$ and 
$S_{\rm dc}^{(1)}(x) =  1/4 ~ \ep^2 (\ln 2 + 1/2) -1/4 ~ \ep^2 \ln \ep$, or 

\begin{equation}
S_{\rm dc}^{(1)}(u) = - \frac{3}{8} (1 - u)\ln(1 - u) + O(1 - u).  \label{eqn_Su1}
\end{equation}

These results have been obtained previously in \citep{mik}, but we have taken the opportunity to correct  
some errors in the expressions for $\lambda$, $u$ and $S_{\rm dc}(x)$ (for $x = 1 - \ep$). The expression given for $S_{\rm dc}^{(1)}(u)$ (here Eq. \ref{eqn_Su1}) was however error free.

\begin{figure}[h]
 \centering
 \begin{minipage}[b]{5.5cm}
 \includegraphics{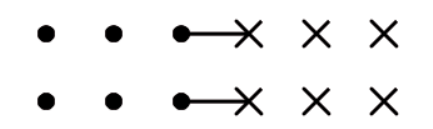}
 \caption{Bond occupation leaving the ground state requires at least two bond sites occupied}
 \label{mika-fig4}
 \end{minipage} \hspace{2cm}
 \begin{minipage}[b]{5.5cm}
 \includegraphics{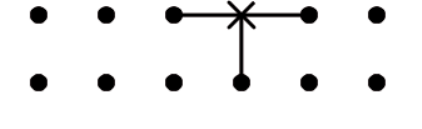}
 \caption{Spin occupation leaving the ground state requires at least three bond sites occupied}
 \label{mika-fig5}
 \end{minipage}
\end{figure}

Remarkably, starting from $u = 1$, Eq. (2.6) gives a lower entropy increase than does Eq. (\ref{eqn_Su1}). The reason is subtle and the explanation goes as follows. 
When leaving the ground state, each Ising system evolves towards states that maximize the entropy. In other words, the new states are such that the energy increase is minimized. In the ground state of the double chain, there are no (energy-)bonds on the bond sites. Considering bond occupation, the smallest energy quantity, that can be added to the chain is achieved by occupying one bond site, which  -  in accordance with the constraint introduced in Section 1  -  requires a second suitable bond site to be occupied as well (see \figurename~\ref{mika-fig4}). 

In contrast, if we consider spin occupation (see \figurename~\ref{mika-fig5}), leaving the ground state would yield a minimum energy increase 
corresponding to the addition of three bonds. Bond occupation is therefore the correct approach to the physics of the double chain near the ground state, because it is the only way to minimize the energy increase and to minimize the entropy increase. 
This is the profound reason why we give so much importance to the bond occupation approach, rather than the spin occupation scenario, 
for the treatment of the behavior of the double chain
in the vicinity of the ground state $u = 1$. Notice however  that this argumentation
holds only for the double chain and not for the full two-dimensional lattice
as treated by Onsager.

Consequently, let us now treat the spin distribution on the double chain, applying
bond occupation near the ferromagnetic ground state $u = 1$. For this purpose,  we must consider all the different possibilities to distribute
the double bond over the double chain. For instance, in \figurename~\ref{mika-fig4}, where the double chain is of length $n = 5$, there are five configurations. 
They each contain an equal number of +1  and -1 spins: the spin system is disordered near $u = 1$. 
This argument holds for any value $n$ of the chain length and one can 
conclude that even for the smallest excitation above the ground state in
the double chain, the spin system is disordered.

\subsection{Open nets}

\begin{figure}[h]
\centering\includegraphics[width=120mm]{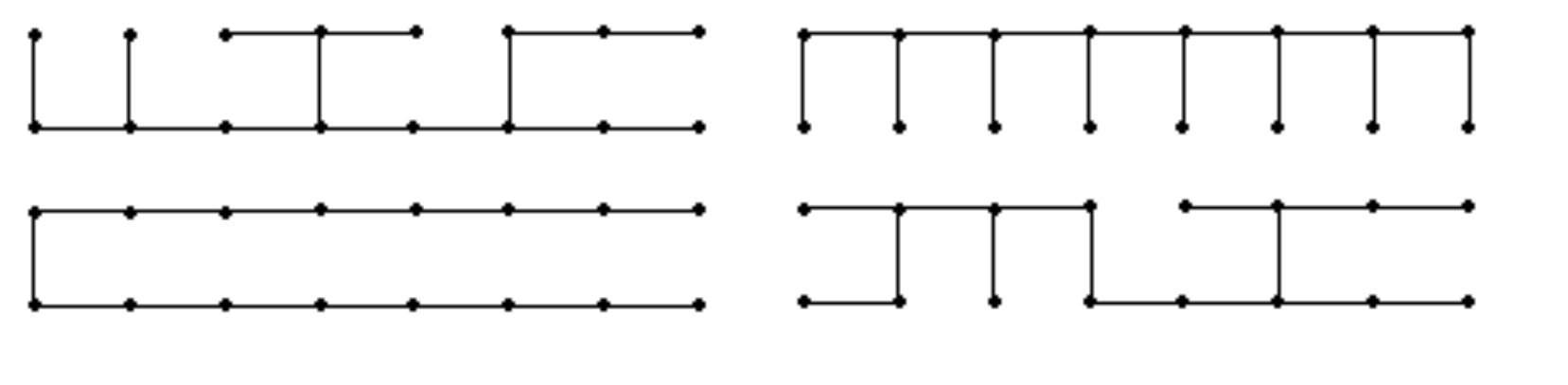} 
\caption{\label{mika-fig6} Four examples of possible open nets on a double chain of length  $n = 7$.}
\end{figure}

An open net on a double chain connects all lattice points and contains no closed loops  (see \figurename~\ref{mika-fig6}). It has length $2n + 1$, the 
number of bond sites on the net. It should be stressed that the open net set is a set of bond sites and not of bonds, in other words, the open net can carry
different numbers of bonds, varying from zero to a maximum number. \figurename~\ref{mika-fig6} shows four examples of open nets for a double chain with $n = 7$ with 15 bond sites each.

\begin{figure}[h]
\centering\includegraphics[width=120mm]{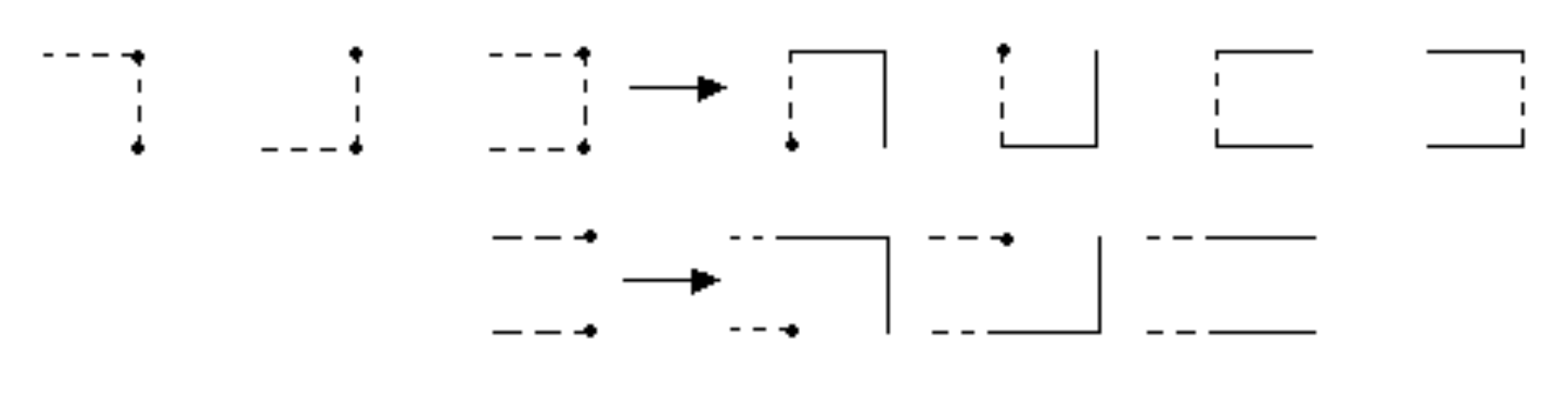} 
\caption{\label{mika-fig7}
Extension of open nets $n$ (dashed lines represent right (last) cell of these nets) to open
              nets  $n + 1$: \newline	
              upper left:     Cells at right end of nets $n$  with vertical bond site at  $n$. \newline
              upper right:   Four possible extensions (to $n + 1$) of nets $n$  with vertical bond site at
                                    $n$. Note: The fourth extension is obtained shifting the vertical bond in
                                    the third extension to the right end. \newline
              lower left:    Cell on right end of nets $n$  without vertical bond site at  $n$. \newline
              lower right:  Three possible extensions (to $n + 1$) of nets $n$  without vertical bond site
                                    at  $n$. Note: These extensions are identical to the first to third 
                                    extension of nets $n$  with vertical bond site at  $n$.}
\end{figure}

We shall have to consider $all$ possible open nets on a given double chain of length $n$. Their number $l(n)$ can be calculated  analytically.
 Let $o_n$ and  $f_n$  be the number of open nets with and without vertical net-bond site at  length $n$.   The sequential construction is schematically represented in \figurename~\ref{mika-fig7}.

\begin{figure}[h]
\centering\includegraphics[width=80mm]{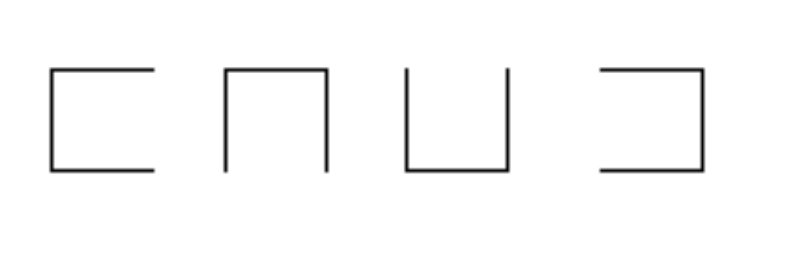} 
\caption{ \label{mika-fig8}
The four open nets on the double chain  of length $n = 1$.}
\end{figure}

We have $o_0 = 1$, $f_0 = 0$  and  $o_1 = 3$, $f_1 = 1$  (\figurename~\ref{mika-fig8}), chain and net for  $n = 0$  being represented by a single vertical bond site.

It is apparent from \figurename~\ref{mika-fig7}, that for $n  >  1$ one has an extension rule encapsulated by 
\begin{eqnarray}
o_{n+1} &=& 3 o_n + 2 f_n,\\ 
f_{n+1} &=& o_n + f_n.
\end{eqnarray}
The eigenvalues of this system are $\lambda_{\pm}= 2 \pm \sqrt{3}$, and   $l(n) = A \lambda_+^n +B \lambda_-^n$
with A, B fixed by  $l(0) = 1$ and $l(1) = 4$.  It follows that 
\begin{eqnarray}
l(n) = \frac{1}{2 \sqrt{3}} [ ( 2 + \sqrt{3})^{n+1} - (2 - \sqrt{3})^{n+1} ] .    \label{eqn_lofn}
\end{eqnarray}
Furthermore, the $l(n)$ obey the recursion relation: $l(n+1) = 4 l(n) - l(n-1)$, their exact values are listed in Table 1 for $n = 0, \ldots,  10$.
We note, that the ratio $l(n+1)/l(n)$ converges rapidly to  $\lambda_+ = 2 + \sqrt{3}\simeq 3.732050808 \ldots$, meaning that $l(n)$ grows as $\lambda_+^n$. For $n \geq 7$,  the ratio agrees with its asymptotic value to seven decimal places or better.

It was shown in \citep{mik} that, near  $u = 0$ and for large $n$, each open net on the Ising systems 
contributes the same amount to $S_b(u)$:
\begin{equation} 
S_b^{(0)}(u)= \ln 2 -\frac{1}{2} \Big(\frac{L}{N}\Big)^2 u^2 ,
\end{equation}                        
$i.e.$ the open net entropy, starting at $u = 0$, decreases faster than does the  -  physically correct  -  "thermodynamic" entropy of Ising systems given in Eq. (\ref{eqn_Su0}). 

We have dealt with this discrepancy in \citep{mik} and for the sake of completeness we briefly describe the procedure.
One calculates  the mean entropy $\overline{S_b}(u)$  from the entropy of all open-net configurations (all possible nets with all possible combinations of bond arrangements on the nets' bond sites), introducing a set of weighting parameters for  each of the net entropies. These 
parameters are  determined in turn by demanding that $\overline{S_b}(u)$  be maximized near $u = 0$. This  yields the entropy 
$\overline{S_b}(u)$  with the correct behaviour (Eq. \ref{eqn_Su0}) near $u = 0$, and  the critical energy $u_c = \sqrt{2/3}$
 for the double chain  from 
$\overline{S_b}(u_c) = 0$. Parenthetically, in \citep{mik} we have analogously treated the periodic square chains with 4, 6, 8, and 10 rows, obtaining for each of these systems the {\em same} entropy function  $\overline{S_b}(u)$, and  -  from $\overline{S_b}(u_c) = 0$  -  the {\em same} critical energy  $u_c = \sqrt{1/2}$,  $i. e.$  the Onsager-Kaufman solution for the infinite two-dimensional Ising square system. 

For the problem at hand, we again consider  all possible open nets and all of their bond sites, but we limit ourselves to occupation of only $one$ 
net-bond site at the time, leaving  $2n$ net-bond sites empty. This means that, for large $n$, $S_b(u)$ is approximately 0, thus avoiding  the necessity of calculating explicitly $S_b(u)$. In compliance with the constraint that each cell of the chain has to have an even number of bonds, 
we can obtain {\em directly} the characteristic energy $u^{(n)}_{ch}$ of a double chain of length  $n$ without calculating the entropy.  
For large $n$, the characteristic  energies  $\{ u^{(n)}_{ch} \}$  converge to the critical energy $u_c = \sqrt{2/3}$ of the infinite double chain, the value found in \citep{mik}.

We note, that for the linear chain and for large  $n$ the ratio $L/N = 1$ in Eqs. (3.4) and (3.9), and that therefore  the discrepancy discussed above does not exist. 

\subsection{Occupation of open nets by one bond}
               
If there is no bond on a net, the not-net-bond sites are also empty. If, however, there is one bond on the net, there has also to be at least one more bond on not-net-bond sites in order to comply with the constraint discussed in Section 1 (for $n = 1$ see Fig. 9). 

\begin{figure}[H]
\centering\includegraphics[width=120mm]{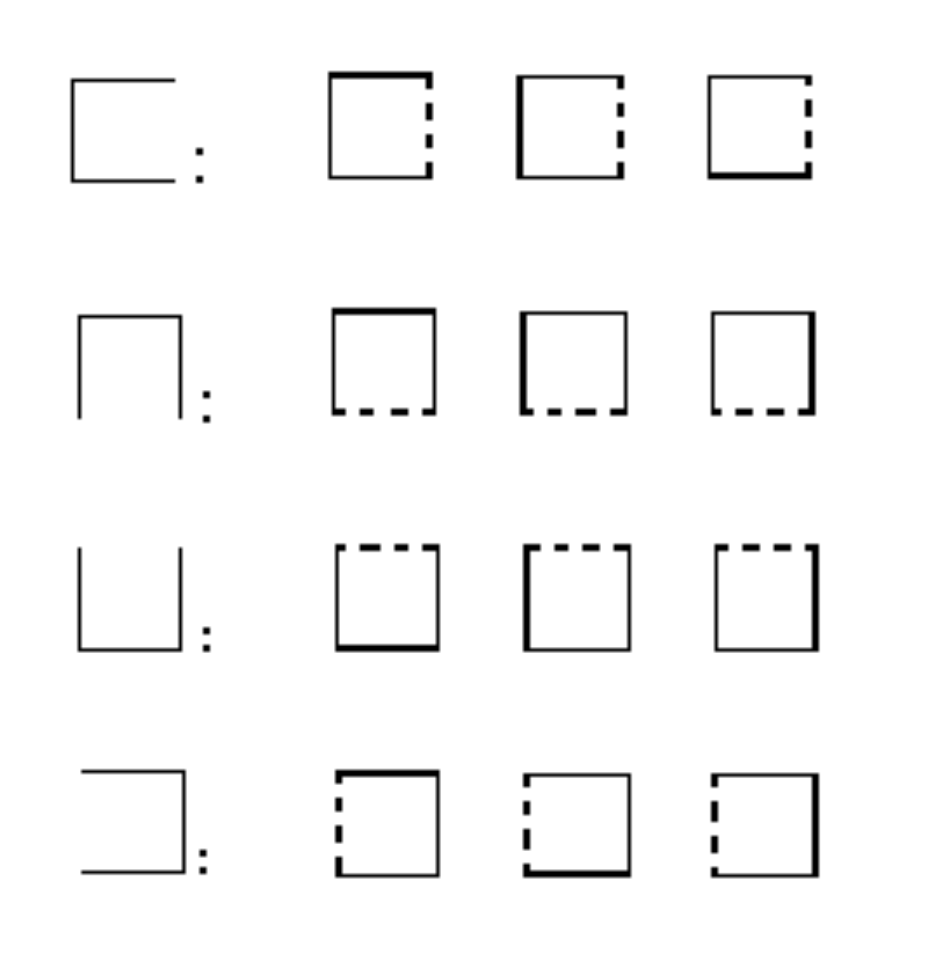} 
\caption{ \label{mika-fig9}
Occupation of the four open nets  $n = 1$  by one bond each. 
             left:   the four nets. 
             right: the three possibilities to place one bond on each net (thick lines). The dashed lines represent the chain-bond sites not lying on the net, which have to be occupied as well, in order to comply with the constraint of an even number of bonds for each cell.}
\end{figure}

To obtain $u^{(n)}_{ch}$, we shall need, for fixed $n$,  all  nets, and we have to take into account occupation of each net-bond site of each of these nets with one  bond at the same time (see \figurename~\ref{mika-fig9} for  $n = 1$). Thus, we get a total number of bonds $i^{(n)}_{\rm tot}$ on net- and on not-net-bond sites. For large  $n$, bonds on net-bond sites become more and more negligible, and the remaining bonds on the not-net-bond sites will finally determine the critical energy.

For chains of arbitrary length $n$, one analyzes the geometry of the open nets and considers the position of the bond site occupied by the one bond on the net. For fixed $n$, one finds  the number of net types and the number of bonds for each type; this makes up the
contribution of each net type to $i^{(n)}_{\rm tot}$. For the sake of brevity, we only present the details of the calculation for the first few 
($n= 0, 1, 2, 3$) values of $i^{(n)}_{\rm tot}$. The procedure is straightforward, but becomes increasingly tedious as the length $n$ of the double chain  continues to grow.

\begin{figure}[h]
\centering\includegraphics[width=160mm]{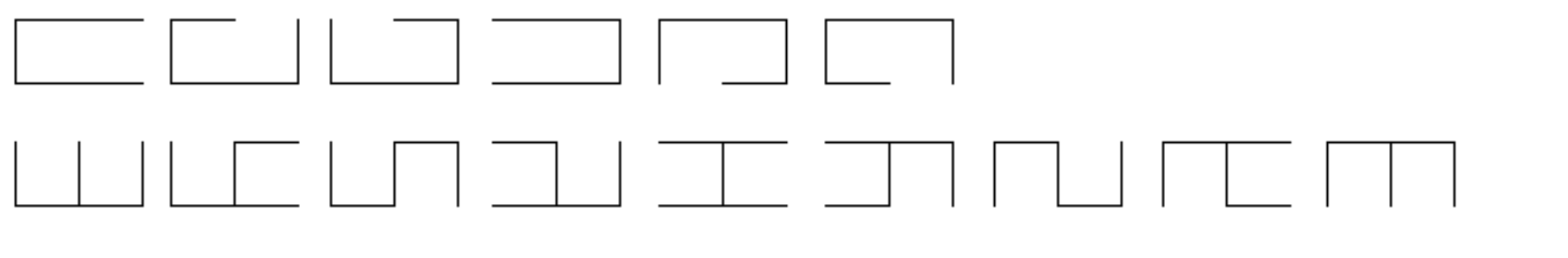} 
\caption{ \label{mika-fig10}
Double chain of length $n = 2$:  6 nets of the first type and 9 nets of the second type.}
\end{figure}

\begin{figure}[h]
\centering\includegraphics[width=120mm]{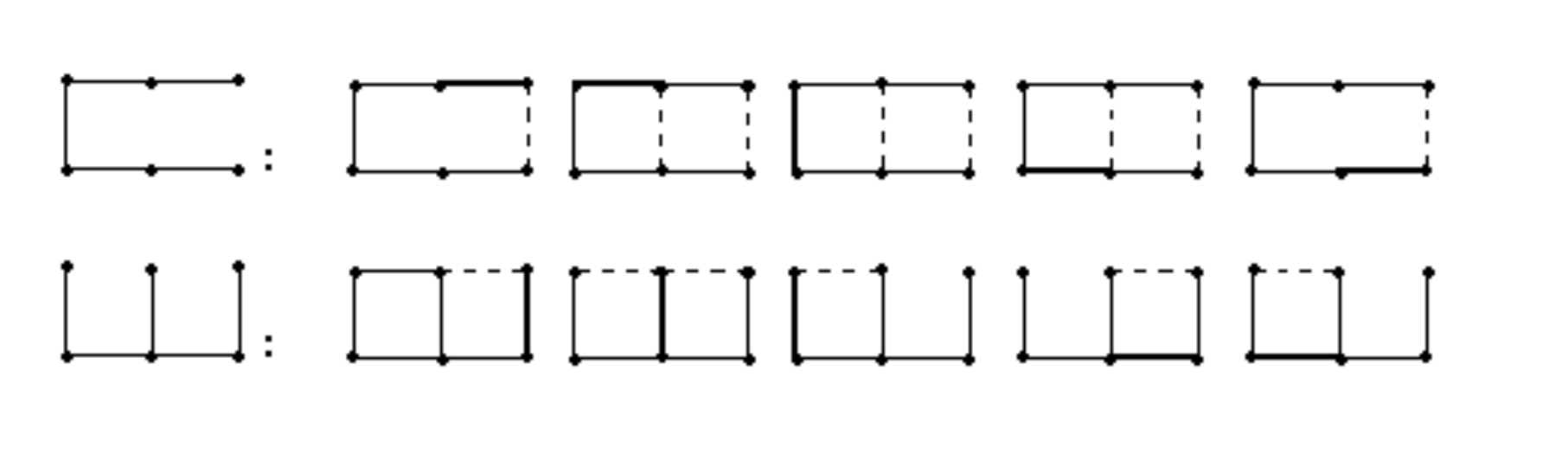} 
\caption{ \label{mika-fig11}
Contribution of one single net ($n =2$)  to $i^{(2)}_{\rm tot}$ (two examples  -  see Fig. 10). 
              upper line: contribution of one of the nets of the first type, counting from 
                                left to right, 2 + 3 + 3 + 3 + 2  = 13 bonds. 
              lower line: contribution of one of the nets of the second type, counting from 
                                left to right, 2 + 3 + 2 + 2 + 2  = 11 bonds.}
\end{figure}

For  $n = 0$,  we have  $i^{(0)}_{\rm tot} = 1$.  For  $n = 1$,  we have (see \figurename~\ref{mika-fig9})  $i^{(1)}_{\rm tot} = 24$.  For  $n = 2$, the 15 nets can be divided into two types:  6 nets of the first type  and 9 nets of the second type  (\figurename~\ref{mika-fig10}).  Each net of 
the first type contributes 13 bonds and each net of the second type contributes 11 bonds to  $i^{(2)}_{\rm tot}$ 
 (\figurename~\ref{mika-fig11}).  The result is $i^{(2)}_{\rm tot} =6 \times 13+9 \times 11= 177$ bonds.

\begin{figure}[h]
\centering\includegraphics[width=120mm]{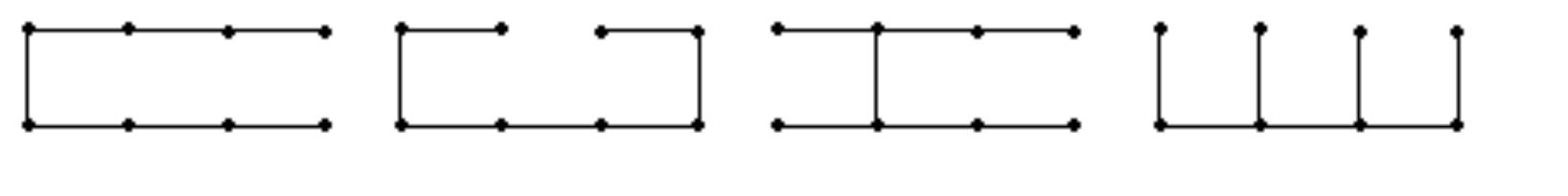} 
\caption{ \label{mika-fig12}
The four types of nets for  a double chain of length $n = 3$. From left to right: 6 nets of the first type contribute 22 bonds each  to  $i^{(3)}_{\rm tot}$, 2 nets of the second type contribute 20 bonds, 30 nets of the third type contribute 18 bonds , and 18 nets of the fourth type contribute 16 bonds. Therefore, 
 $i^{(3)}_{\rm tot} =6 \times 22 +2 \times 20 +30 \times 18 +18 \times 16  =1000 $ bonds.}
\end{figure}

Similarly, for  $n = 3$, we find four net types (see \figurename~\ref{mika-fig12}) and  $i^{(3)}_{\rm tot}$ = 1000. 
The  numbers $i^{(n)}_{\rm tot}$ obtained for  $n = 4, \ldots , 10$  are given in Table 1.

\subsection{The critical energy}  

As a first step, we introduce the mean number  $i(n)$  of bonds (bonds on not-net-bond sites plus those occupying the one net-bond site) per net-bond site. With the number  $l(n)$ of open nets and with the length  $2n +1$  of each net we have
\begin{equation}       
i(n) = \frac{  {i^{(\rm n)}_{\rm tot}}   }      {l(n)(2n+1)} . \label{eqn_iofn}
\end{equation}
These numbers  are listed in Table 1 for the first 10 values of $n$. 
 
\begin{figure}[h]
\centering\includegraphics[width=80mm]{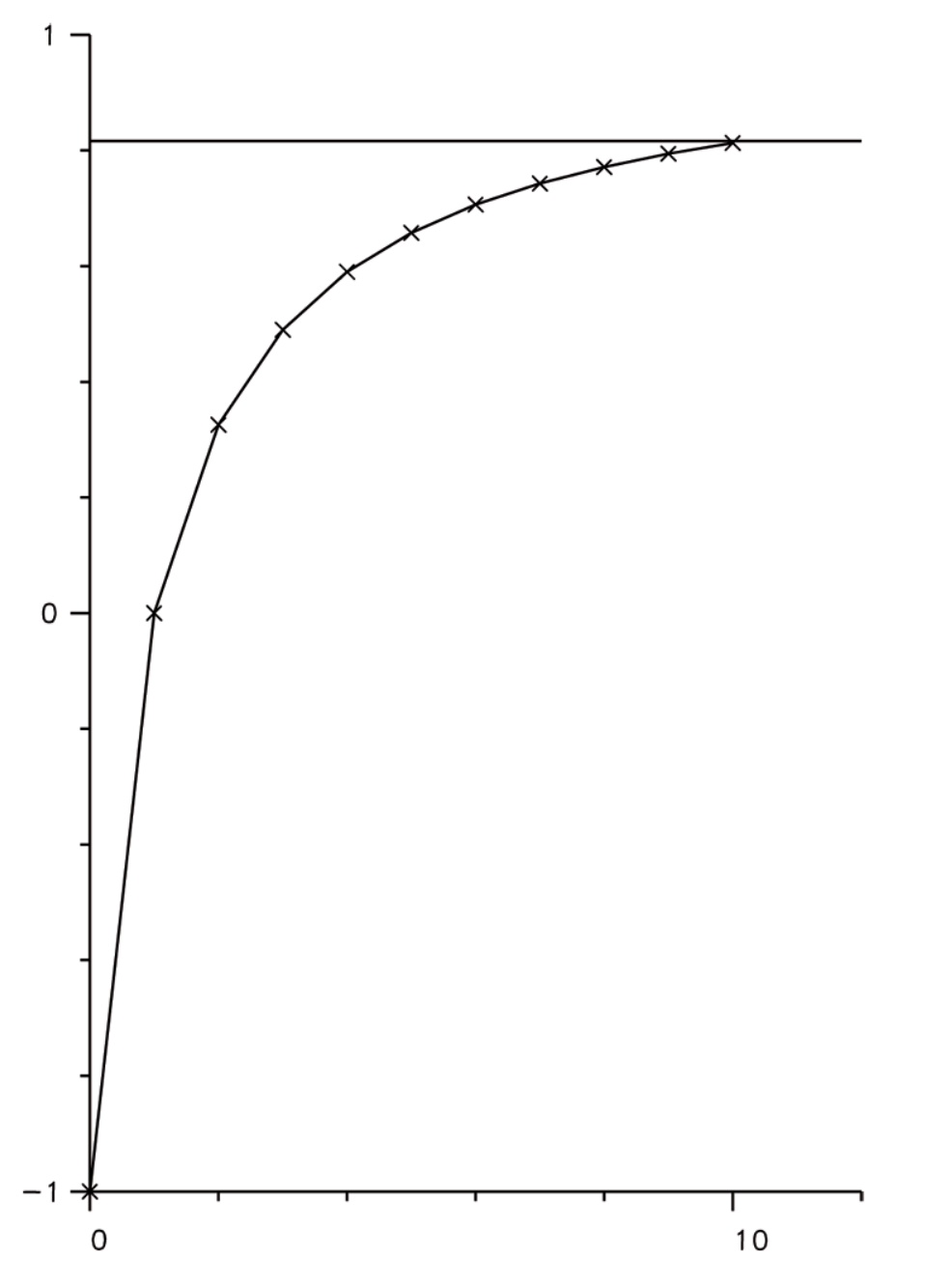} 
\caption{ \label{mika-fig13}
The characteristic energies  $u^{(n)}_{ch}$  for $n = 0, 1, \ldots , 10$  converging to the critical 
               energy $u_c = \sqrt{2/3} \simeq 0.816496 \ldots$ of the infinite double chain.}
\end{figure}

With the number  $L =  3n + 1$ of  bond sites on the chain  and  
$u^{(n)}_{ch}  =  1 - 2i(n)/L$   (Eq. \ref{eqn_uofn}), we get characteristic energies  $u^{(n)}_{ch}$ shown in Table 1 and in \figurename~\ref{mika-fig13}. 
For sufficiently large $n$, $u^{(n)}_{ch}$ converges nicely to the critical energy $u_c = \sqrt{2/3}$ of the infinite double chain.

For a precise analysis of the convergence as a function of $n$, it could be useful to include values  $i^{(n)}_{\rm tot}$ for  $n$ > 10.  
However, given the rapid convergence displayed in Table 1 (at $n=10$, the characteristic energy is within a fraction of a per cent to its asymptotic value),  the cumbersome task of calculating exactly $i^{(n)}_{\rm tot}$ may not be worth the effort. A few scaling rules are easily obtained from 
our analysis: $l(n) \propto \lambda_+^n$, $i^{(n)}_{\rm tot} \propto n^2 \lambda_+^n$, and $i(n) \propto n$.

\begin{table}[t]
\begin{center}
\begin{tabular}{r|r|r|r|r|}
$n$ & $l(n) $& $ i^{(n)}_{\rm tot}$ & $i{(n)}$ & $u^{(n)}_{ch}$ \\ \hline
0&1& 1& 1& -1 \\
1&4& 24& 2& 0 \\
2&15& 177& 2.36& 0.3257 \\ 
3&56& 1 000& 2.5510& 0.4898 \\
4& 209& 5 013 &2.6651& 0.5900 \\
5& 780& 23 504& 2.7304& 0.6577 \\
6& 2 911& 105 629& 2.7912& 0.7062 \\
7& 10 864& 461 072& 2.8294& 0.7428 \\
8& 40 545& 1 970 281& 2.8585& 0.7713 \\
9& 151 316& 8 281 600& 2.8806& 0.7942 \\
10& 564 719& 34 399 495& 2.9007& 0.8129 \\
\end{tabular}
\caption{The exact results,  $l(n)$ (Eq.\ref{eqn_lofn}) and  $i^{(n)}_{\rm tot}$; the derived results, 
               $i(n)$ (Eq.\ref{eqn_iofn}) and $u^{(n)}_{ch}$ (Eq.\ref{eqn_uofn}) for $n = 0, 1,  \ldots , 10$.}
\end{center}
\end{table}

\section {Conclusions} 

The basic idea of this work is to use energy and entropy to describe the Ising systems, applying a general suggestion of Gibbs dating 
back to 1875. 
For the Ising double chain of length $n$, we obtain analytically the value of the characteristic energy  $\{ u^{(n)}_{ch} \}$
which serves as a good approximation to the critical energy $u_c$ with increasing $n$. The special status of the critical energies 
is exemplified in Appendix A. 
Our approach uses a bond occupation framework which not only describes the degeneracy of the linear chain, but also of all other Ising systems. 
The origin of the physical mechanism of melting arises from the presence of a critical energy $u_c$ away from the ground state energy  $u = 1$, The essential point is, that already one bond  on the bond sites of an open net produces bonds on not-net-bond sites, leading to a deeper understanding of melting: The critical energy is the minimum averaged amount of energy, that has to be added to the chain, if only one of its -- randomly chosen  -- bond sites is to be occupied.

On the basis of our special type of microcanonical description, further research concerning phase transition and critical energy should be possible, 
for instance the investigation of other (e.g. 3-dimensional) Ising arrangements like the periodic square chain with four rows ("double ladder"), and possibly the description of the influence of the geometric parameters of these arrangements on their physical properties, such as the critical energy.

\section*{Acknowledgments}

The author wishes to thank Klaus Bartels for his continuous support and his valuable contributions in obtaining  $i^{(n)}_{\rm tot}$. He also thanks Marvin Goblet, who first determined numerically the number of nets $l(n)$ , as well as Kurt Wingerath and Christine Kaiser for their assistance in the preparation of the manuscript.

\newpage
\section*{Appendix A. The critical energy of the kagom{\'e} net}

The particular role of the critical energy can be appreciated with a look at he known two-dimensional Ising lattices.
The four classical Ising systems (Onsager (ON) \citep{ons,kau}, triangular (TR) \citep{new,wan,hou,tem,hus},
honeycomb (HC) \citep{hus,syo}, kagom{\'e} (KG) \citep{kan}) have the following critical energies \citep{dom}, normalized in \citep{mik}:
\begin{equation}
u_c^{\rm ON} = \frac{1}{\sqrt{2}}, \quad u_c^{\rm TR}= \frac{2}{3},  \quad u_c^{\rm HC} = \frac{4}{3\sqrt{3}} \quad 
u_c^{\rm KG} = \frac{1}{6} + \frac{1}{\sqrt{3}}. \tag{A.1}
\end{equation}
We will consider only the kagom{\'e} lattice.

\begin{figure}[H]
\centering\includegraphics[width=80mm]{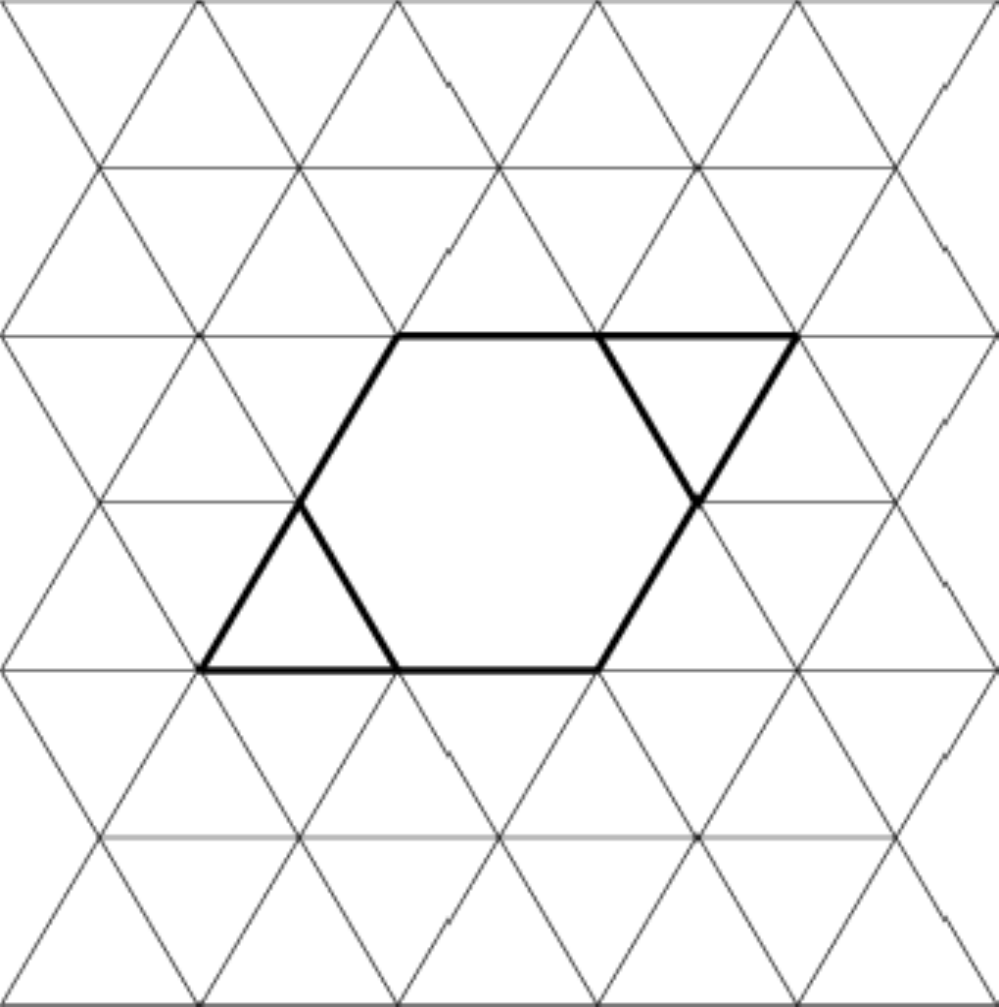} 
\caption{\label{mika-fig14} The unit cell of the kagom{\'e} lattice. It can be divided into eight triangles, two belonging to the triangular 
lattice and six (not shown) to the honeycomb lattice.}
\end{figure}

The unit cell of the kagom{\'e} net has six bond sites (\figurename~\ref{mika-fig14}). All six bond sites cannot be occupied and, since
$i$ must be even, only four bond sites per unit cell can at most be occupied, leading to an antiferromagnetic 
ground state at $u = -1/3$. The unit cell can be divided into eight triangles, two belonging to the triangular
and six (not shown) to the hexagonal net. The critical energy therefore satisfies:
\begin{equation}
u_c^{\rm KG} = \frac{2}{8}u_c^{\rm TR} + \frac{6}{8}u_c^{\rm HC}. \tag{A.2}
\end{equation}
\noindent Corresponding results for $u_c^{\rm TR}$ and $u_c^{\rm HC}$ are still lacking.

\end{document}